\documentclass[twocolumn,superscriptaddress,
 amsmath,amssymb,
 aps,
 prl,
]{revtex4-1}
\usepackage{dcolumn}
\usepackage{bm}
\usepackage{graphicx}
\usepackage{comment}
\usepackage[caption=false]{subfig}
\usepackage{color}

\newcommand{\bleu}{\textcolor{black}}
\newcommand{\blue}{\textcolor{black}}
\newcommand{\newblue}{\textcolor{black}}

\newcommand{\bea}{\begin{eqnarray}}

\newcommand{\eea}{\end{eqnarray}}

\begin{document}

\title{\bleu{Extreme statistics and spacing distribution in a Brownian gas correlated by resetting}}
\author{Marco Biroli}
\affiliation{LPTMS, CNRS, Univ.  Paris-Sud,  Universit\'e Paris-Saclay,  91405 Orsay,  France}
\author{Hernan Larralde}
\affiliation{Instituto de Ciencias Fisicas, UNAM, Av. Universidad s/n, CP 62210 Cuernavaca Morelos, Mexico}
\author{Satya N. Majumdar}
\affiliation{LPTMS, CNRS, Univ.  Paris-Sud,  Universit\'e Paris-Saclay,  91405 Orsay,  France}
\author{Gr\'egory Schehr}
\affiliation{Sorbonne Universit\'e, Laboratoire de Physique Th\'eorique et Hautes Energies, CNRS UMR 7589, 4 Place Jussieu, 75252 Paris Cedex 05, France}


\begin{abstract}
We study a one-dimensional gas of $N$ Brownian particles that diffuse independently, but are {\it simultaneously} reset to the origin at a constant
rate $r$. The system approaches a non-equilibrium stationary state (NESS) with long-range interactions induced by the simultaneous resetting. Despite the presence of strong correlations, we show that several observables can be computed exactly, which include the global average density, the distribution of the position of the $k$-th rightmost particle and the spacing distribution between two successive particles. Our analytical results are confirmed by numerical simulations. We also discuss a possible experimental realisation of this resetting gas using optical traps.   
\end{abstract}

\maketitle


\newpage

While the properties of a gas of noninteracting particles are well understood, those of an interacting gas, in particular in the presence
of a long-range interaction between particles, are much less so. A notable exception is the celebrated Dyson log-gas in one-dimension, that appears in the spectral statistics of random matrix theory (RMT). Indeed, the statistics of the eigenvalues of Gaussian random matrices play a major role in several areas of science,
from nuclear physics, quantum chaos, mesoscopic transport, all the way to finance and information theory \cite{mehta,forrester,vivo,bouchaud}. 
For an $N \times N$ matrix (real symmetric, complex Hermitian or quaternionic symplectic) 
with independent Gaussian entries, the joint probability distribution function (JPDF) of the $N$ real eigenvalues $\{x_i\}$  can be expressed as a Boltzmann weight 
$P[\{x_i \}] \propto \exp(- \beta E[\{x_i \}])$ with the energy given by $E[\{x_i \}] = \frac{1}{2} \sum_{i=1}^N x_i^2 - \frac{1}{2}\sum_{i\neq j} \ln |x_i - x_j|$,
where the Dyson index $\beta  = 1, 2, 4$ corresponds to the three symmetry classes mentioned above \cite{mehta,forrester}. Thus, the eigenvalues $x_i$ can be interpreted as the positions of $N$ particles on a line in the presence of a confining harmonic potential, with pairwise logarithmic repulsion between them. This is Dyson's log-gas~\cite{Dyson_62}, which has been a fundamental cornerstone \cite{forrester} in 
understanding the role of strong correlations on several spectral observables such as the average density of eigenvalues~\cite{Wigner_58}, 
the largest eigenvalue~\cite{TW_94,TW_96,BDJ_1999,MS_14} (i.e., the position of the rightmost particle in the gas) and 
the spacing distribution between successive eigenvalues~\cite{Wigner_51,mehta,forrester}.  
These observables can be computed exactly for the log-gas, thanks to a special analytical structure of the particular form of the JPDF~\cite{mehta,forrester}. Moreover, they have been measured experimentally in a variety of systems, from nuclear physics and quantum chaos~\cite{WM_09} to liquid crystals~\cite{TS_10} and fiber lasers~\cite{Davidson_12}. Unfortunately, there exist very few long-ranged \bleu{correlated} gases, even in one-dimension, for which these observables can be computed, with perhaps the exception of the $1d$-jellium model \bleu{where the pairwise repulsion is linear}~\cite{lenard_61,prager_62,baxter_63,DKMS_17,DKMS_18,CGJ_22,FMS_22}.
\begin{figure}[ht]
\includegraphics[width  = 1.\linewidth]{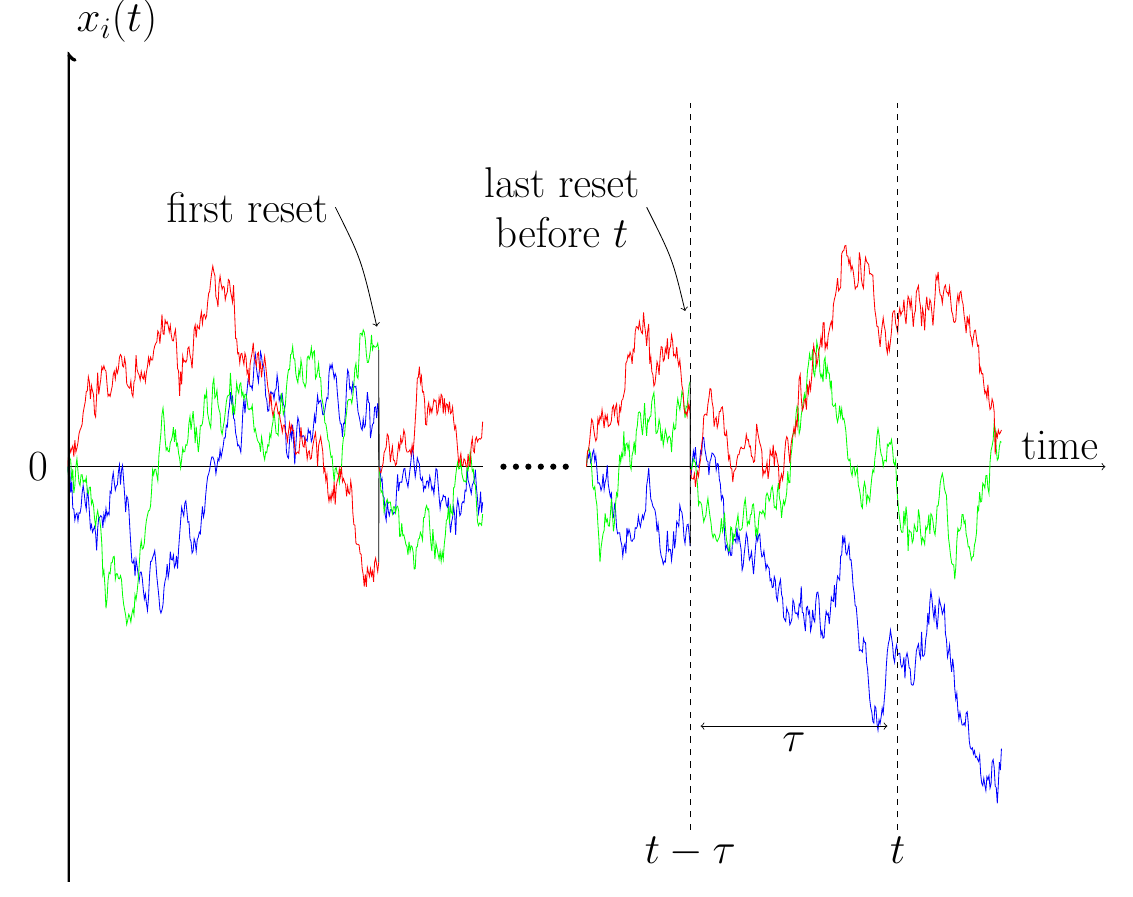}
\caption{\bleu{Schematic trajectories of $N=3$ Brownian motions 
undergoing simultaneous resetting to the origin at random times. 
The observation time is marked by $t$ and the time of the last 
reset before $t$ is marked by $t-\tau$. During the last period $\tau$, 
the particles evolve independently as free Brownian motions.}}\label{Fig_intro_resetting}
\end{figure}

It is therefore natural to look for other experimentally realisable long-ranged \bleu{correlated} particle systems for which these observables can be computed analytically. Motivated by the recent theoretical and experimental advances in the field of stochastic resetting~\cite{EMS_20,PKR_22,Gupta22,Bertin}, in this Letter we propose a new \bleu{many-particle model that, despite the presence of strong correlations induced by dynamics, is solvable for all the spectral observables mentioned above}. 

A single particle subjected to stochastic resetting has been studied extensively over the last decade~\cite{EM_11,EM_11b,EM_14,KMSS_14,MSS_15,PKE_16,Reu_16,MV_16,PR_17,BEM_17,CS_18,Bres_20,Pinsky_20,TPSRR_20,BBPMC_20,FBPCM_21,BMS_22}. Consider, for simplicity, a single Brownian particle diffusing on a line with diffusion constant $D$, starting at the origin. With rate $r$, the particle's position is reset back to the origin and the free diffusion restarts. This resetting move breaks detailed balance and drives the system into a non-equilibrium stationary state (NESS) where the position distribution becomes non-Gaussian~\cite{EM_11,EM_11b}
\bea \label{pstat_reset}
P_{\rm stat}(x) = \frac{1}{2} \sqrt{\frac{r}{D}} \, e^{-|x| \sqrt{\frac{r}{D}}} \;.
\eea
This simple analytical prediction has been verified in recent experiments using holographic optical tweezers \cite{TPSRR_20}. In this Letter, we consider $N$ independent Brownian particles on a line, all starting at the origin, that are
{\it simultaneously} reset to the origin with rate $r$ (this is different from independently reset Brownian particles studied before \cite{EM_11, VAM_22}). This {\it simultaneous} resetting makes the system strongly correlated, and this correlation persists even in the resulting many-body NESS at long times. \bleu{To see this, let us first compute the 
joint distribution $P_r[\{ x_i\},t]$ \newblue{of the positions $x_i$ of the particles at time $t$} (all starting at the origin), where the subscript $r$ denotes the resetting with constant rate $r$. For $r=0$, the particles evolve as $N$ independent Brownian motions and their joint distribution just becomes a product of $N$ independent Gaussians, given by
\bea \label{P0}
P_0[\{ x_i\},t] = \prod_{i=1}^N \frac{1}{\sqrt{4 \pi D t}} e^{- \frac{x_i^2}{4 D t}} \;.
\eea
To see how a nonzero $r$ makes the particles correlated, we proceed as follows. We consider the interval $[0,t]$ and see how many resetting events occur in that interval. With a probability $e^{-rt}$ there will be no resetting in $[0,t]$ -- in that case, the joint distribution at time $t$ will be simply $P_0[\{ x_i\},t]\, e^{-rt}$. When there is at least one resetting event in $[0,t]$, we remark that the state of the system at time $t$ depends only on the time elapsed since the last resetting before $t$. This is because every resetting event brings back all the particles to the origin and hence we only need to keep track of the time since the last resetting. This idea is illustrated in Fig. \ref{Fig_intro_resetting} where $t$ is the observation time and $t - \tau$ is the time at which the last resetting occurs before $t$. Since the evolution between $t-\tau$ and $t$ is free (i.e., without resetting), clearly the joint distribution of the positions at time $t$ is simply $P_0[\{ x_i\},\tau]$. However, $\tau$ itself is a random variable, with a probability density $r\,e^{-r \tau}$ and $\tau$ can vary from $0$ to $t$. Hence we need to multiply $P_0[\{ x_i\},\tau]$ by $r\,e^{-r \tau}\,d\tau$ and integrate $\tau$ from $0$ to $t$. Adding these two contributions, i.e., no-resetting event and the multiple resettings, we get the joint distribution at time $t$ as
\begin{equation}  \label{Pr}
P_r[\{ x_i\},t] = e^{-r t} P_0[\{ x_i\},t]  + r \int_0^t d\tau e^{-r\tau} P_0[\{ x_i\},\tau] \,.
\end{equation}
}
\begin{figure}
\includegraphics[width  = \linewidth]{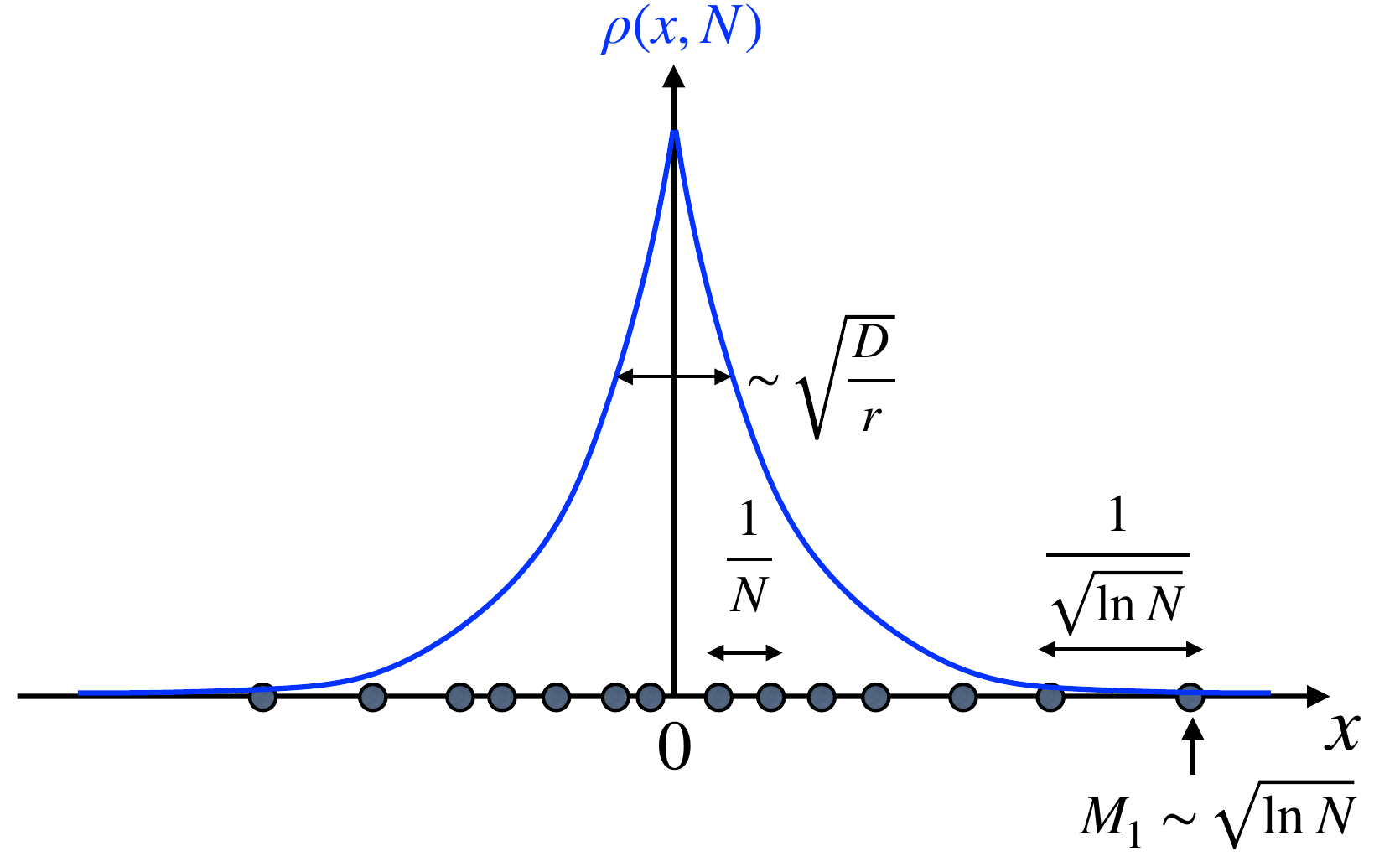}
\caption{The solid blue line shows the average density $\rho(x,N) = \sqrt{\frac{r}{4D}}e^{- \sqrt{r/D}|x|}$. The positions of the particles in a typical sample are shown schematically on the line with most particles living over a distance $\sqrt{D/r}$ around the origin. The typical spacing in the bulk $\sim 1/N$, while it is of order $\sim 1/\sqrt{\ln N}$ near the extreme edges of the sample. The typical position of the rightmost particle $M_1 \sim \sqrt{\ln N}$ for large $N$.}\label{Fig_intro}
\end{figure}
In the long-time limit, the first term in (\ref{Pr}) drops out and we obtain the exact JPDF in the stationary state 
\bea \label{Pr_NESS}
P_{\rm stat}[\{ x_i\}]  = r \int_0^\infty d\tau e^{-r\tau}  \prod_{i=1}^N \frac{1}{\sqrt{4 \pi D \tau}} e^{- \frac{x_i^2}{4 D \tau}} \;.
\eea  
\bleu{This is one of our main results, which merits a few remarks. 
We note that the joint distribution in the stationary state does not 
factorize (even though the integrand inside the integral has a factorized 
form), indicating that the particles are correlated in the steady state. 
The physical origin of these correlations can be traced back to the fact 
that, via simultaneous resetting, the particles are pushed together 
towards the origin, which creates an effective attraction between the 
particles. Note that these correlations or the effective interactions 
between particles in the steady state have a purely {\it dynamical} 
origin and are not inherent interactions between particles as 
in Dyson's log-gas or in the $1d$ jellium model.}
The integral in~(\ref{Pr_NESS}) can, in fact, be performed explicitly 
\bea \label{Pr_NESS_2}
P_{\rm stat}[\{ x_i\}] = \left( \frac{r}{2 \pi D}\right)^{\frac{N}{2}} R_N^{\frac{2-N}{2}} K_{\frac{N}{2}-1} \left( R_N\right) \;,
\eea
where $R_N = \sqrt{\frac{r}{D}}\sqrt{x_1^2 + \cdots + x_N^2}$ and $K_\nu(z)$ is the modified Bessel function of index $\nu$. This makes the correlated nature of the gas manifest, since the JPDF does not factorize, though unlike the log-gas the correlation
is not pairwise but rather ``all-to-all''.  Finally, to see that this resetting gas indeed has long range correlations, we compute the two-point correlations from the JPDF in Eq. (\ref{Pr_NESS}). Noting that $\langle x_i x_j \rangle - \langle x_i \rangle  \langle x_j \rangle =0$ (for $i \neq j$) trivially, the first non-trivial correlator is given by  
\bea \label{correl}
\langle x_i^2 x_j^2 \rangle - \langle x_i^2 \rangle  \langle x_j^2 \rangle = \frac{4D^2}{r^2}
\qquad \forall ~i,j\;,
\eea
which manifestly demonstrates the long-range correlations.
 
\bleu{Given the JPDF in Eq. (\ref{Pr_NESS}), our goal, motivated by the studies in the Dyson log-gas, is to compute three natural 
observables, namely: (i) the average density, (ii) extreme statistics and (iii) the spacing distribution between consecutive particles.}
The reason why these observables can be computed exactly can be seen in the structure of the JPDF in Eq. (\ref{Pr_NESS}), where the
integrand (modulo $e^{-r\tau}$) just corresponds to a set of $N$ independent and Gaussian distributed random variables, parametrised by $\tau$. For a fixed $\tau$, we first compute the statistics of these observables for $N$ independent \blue{and identically distributed (IID)} Gaussian random variables and then integrate over $\tau$. We will see that this simple mechanism leads to rather rich and interesting behaviors of these observables. 

We start with the first basic observable, namely the average density of particles in the stationary state, defined
by $\rho(x,N) = \frac{1}{N}\langle \sum_{i=1}^N \delta(x-x_i) \rangle$, where $\langle \cdots \rangle$ denotes the average over the stationary measure in (\ref{Pr_NESS}). The density $\rho(x,N)$ is normalised to unity and measures the average fraction of particles in $[x,x+dx]$. 
\newblue{Using the invariance of the JPDF in (\ref{Pr_NESS}) under exchange of $i$
and $j$, one sees that $\rho(x,N)$ is also the one-point function $\rho(x,N) = 
\int_{-\infty}^\infty dx_2 \cdots dx_N P_{\rm stat}(x,x_2,\cdots, x_N)$. 
Then, given the factorisation property in Eq. (\ref{Pr_NESS}), we find that
$\rho(x,N)$ coincides with the position distribution $P_{\rm stat}(x)$ of a single particle given in Eq. (\ref{pstat_reset}) and plotted in Fig.~\ref{Fig_intro}.} \newblue{However, this does not mean that the particles are uncorrelated, as seen from the fact the JPDF in Eq. (\ref{Pr_NESS}) does not factorise.} Thus, $\rho(x,N)$ is independent
of $N$ and is supported over the full line. \bleu{This is in contrast with other models with long-range pairwise repulsion, such as the Dyson log-gas and the $1d$ jellium model, where the average density is supported over a finite interval. In the former case, it is the celebrated Wigner semi-circular law~\cite{Wigner_58} while, for the jellium, the average density is flat over a finite interval~\cite{lenard_61,prager_62,baxter_63,DKMS_17}}. 

Moreover, from Eq. (\ref{pstat_reset}), one sees that the density decreases exponentially over a length scale $\sqrt{D/r}$ where most particles are concentrated in a typical sample (see Fig. \ref{Fig_intro}). Hence the typical spacing between particles in the bulk scales as $\sim O(1/N)$ for large $N$. While the average density extends over the full space, in a typical sample, the rightmost (or leftmost) particle is located at a distance of order $O(\sqrt{\ln N})$ from the center \blue{(see later)}. In addition, the spacing between two particles near these extremes scales as $1/\sqrt{\ln N} \gg 1/N$. Thus in a typical sample the gas is denser near the center and sparser near the extremes, as illustrated in Fig. \ref{Fig_intro}.

\begin{figure}[t]
\includegraphics[width = \linewidth]{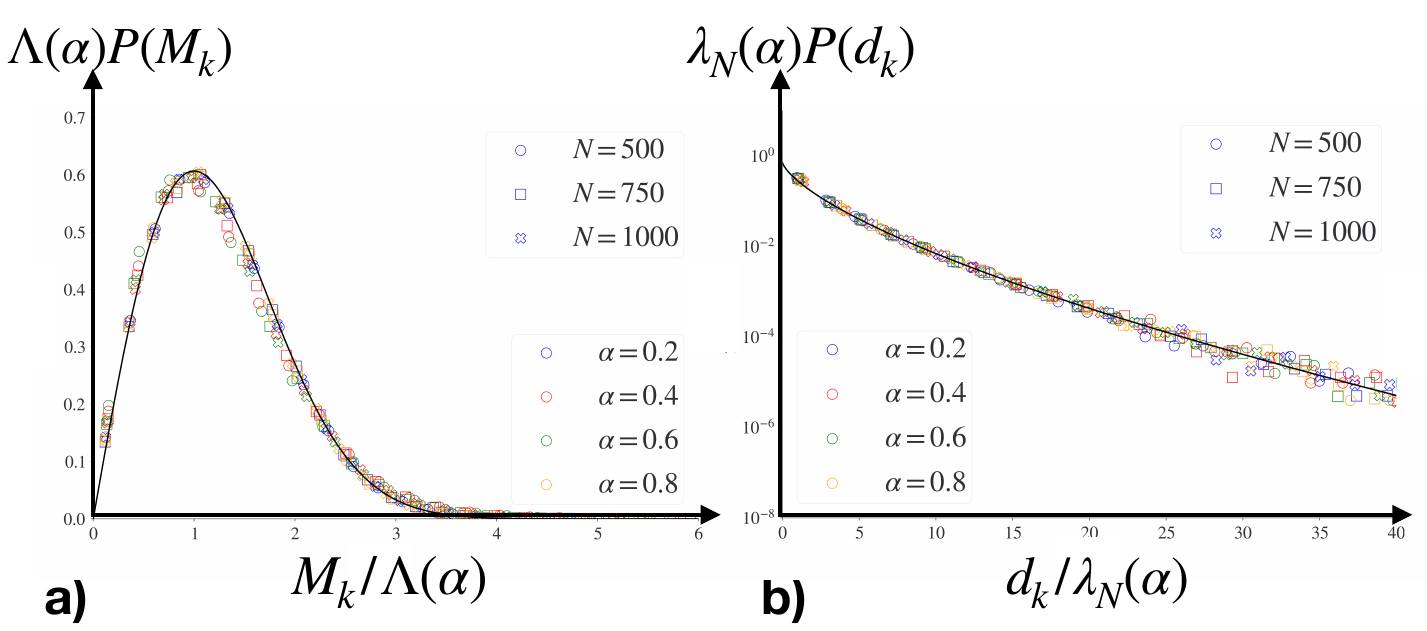}
\caption{{\bf a)} Scaled distribution of the position $M_k$ of the
$k$-th particle from the right: 
$P(M_k) \approx \Lambda^{-1}(\alpha) f(M_k\Lambda^{-1}(\alpha))$ 
with $\Lambda(\alpha)$ given below Eq. (\ref{scaling_kth_max}). 
The symbols represent the results of simulations, 
while the solid curve shows the scaling function $f(z)$ 
in Eq. (\ref{scaling_kth_max}). {\bf b)} Scaled distribution of the 
gap $d_k = M_k - M_{k+1}$ between the $k$-th and the $(k+1)$-th 
particle counted from the right: numerical simulations are 
in perfect agreement with the analytical scaling function $h(z)$ in Eq. 
(\ref{h_of_z}). \blue{We used the parameter values $D=0.5$ and $r=1$}.}\label{Fig_combined}
\end{figure}

Having computed the global density, we now probe the gas at a local 
level by studying the statistics of the positions of individual 
particles and the spacing between them. For this, it is convenient to 
first order the positions $\{x_1, x_2,\cdots, x_N \}$ and label them as 
$\{M_1 > M_2 > \cdots > M_N\}$ where $M_k$ denotes the position of the 
$k$-th particle counted from the right. Thus $M_1 = \max\{x_1, x_2, 
\cdots, x_N \}$ denotes the global maximum, i.e., the position of the 
rightmost particle. \bleu{This observable $M_1$ is well studied when the 
underlying random variables $x_i$ are uncorrelated and its 
distribution is known to belong to the three famous universality 
classes, namely Gumbel, Fr\'echet and Weibull depending on the tails of 
the distribution of $x_i$~\cite{Gum_58,david,SM_14,MPS_20}. 
There has been a lot of interest in 
computing the distribution of $M_1$ in the case where the random 
variables $x_i$ are strongly correlated and very few results are known 
in that case \cite{MPS_20}. One well known example corresponds to the 
Dyson log-gas, where $M_1$ represents the largest eigenvalue of a 
Gaussian random matrix. In this case, the distribution of $M_1$, 
appropriately centered and scaled, follows the celebrated Tracy-Widom 
distribution \cite{TW_94,TW_96,BDJ_1999,MS_14}. Another solvable example 
corresponds to the $1d$ jellium model where the distribution is known to 
be different from the Tracy-Widom law \cite{DKMS_17,DKMS_18}. Similarly, 
the statistics of the $k$-th maximum have been studied for Dyson's 
log-gas \cite{TW_94,TW_96}. One of the main results of this Letter is to 
compute exactly the distribution of $M_k$ for all $k$ in the correlated 
resetting gas. Notably, for $k=1$, we find a new extreme value 
distribution, which is different from the ones mentioned above.}

\bleu{We start by computing the PDF of $M_k$, i.e., the $k$-th maximum 
of the ordered positions $x_i$ \newblue{that are distributed} via the JPDF $P_{\rm 
stat}[\{x_i\}]$ in Eq. (\ref{Pr_NESS}). As for the JPDF, it is 
convenient to exploit the renewal structure in Eq. (\ref{Pr}), also 
depicted graphically in Fig. \ref{Fig_intro_resetting}. It is clear, 
then, that in the stationary state ($t \to \infty$ limit), the PDF of 
$M_k$ can be expressed as
\bea \label{pdf_kth_max}
{\rm Prob.}(M_k = w) = r \int_0^\infty d\tau e^{-r\tau}\, {\rm Prob.}(M_k(\tau) = w) \:,
\eea
where $M_k(\tau)$ is the $k$-th maximum of a set of $N$ independent 
Brownian motions each of duration $\tau$, i.e., drawn from the Gaussian 
distribution \newblue{$\exp{[-{x_i^2}/{(4 D 
\tau})]}/{\sqrt{4 \pi D \tau}}$}. The distribution of the $k$-th maximum of $N$ IID Gaussian 
random variables is well studied in the literature and is reproduced in 
the \newblue{Supplemental Material} \cite{SM}. Here we just state the main results. We set $k 
= \alpha N$ and take the limit of large $N$, keeping $0<\alpha < 1$ 
fixed. In this limit, the distribution of $M_k(\tau)$ approaches a 
Gaussian form with mean $w^* = \sqrt{4 D \tau}\,{\rm erfc}^{-1}(2 
\alpha)$ and variance $\propto 1/N$ (here ${\rm erfc}^{-1}(z)$ is the inverse of the complementary
error function ${\rm 
erfc}(z)=(2/\sqrt{\pi})\, \int_z^{\infty} e^{-u^2}\, du$). In the large $N$ limit, the 
distribution of $M_k(\tau)$ essentially approaches a $\delta$-function 
centred at $w^*$, i.e., $ {\rm Prob.}(M_k(\tau) = w) \to \delta(w - 
\sqrt{4 D \tau}\,{\rm erfc}^{-1}(2 \alpha))$. Substituting this behavior 
in Eq. (\ref{pdf_kth_max}) we arrive at
\bea \label{scaling_kth_max}
{\rm Prob.}(M_k = w) \approx \frac{1}{\Lambda(\alpha)} \, f\left( \frac{w}{\Lambda(\alpha)}\right) \;, \; f(z) = 2 z e^{-z^2} 
 \eea 
 with $z \geq 0$ and $\Lambda(\alpha) = \sqrt{4D/r}\,{\rm erfc}^{-1}(2 \alpha)$. In the large $N$ limit, the scaling function $f(z)$ is thus supported only over $z \geq 0$ and is universal, i.e., it is independent of $\alpha$. For $\alpha = O(1)$, this gives us the behavior for the $k$-th maximum in the bulk, while setting $\alpha = k/N$ with $k = O(1)$ we can probe the $k$-th maximum near the global maximum $M_1$. In this limit, using ${\rm erfc}^{-1}(2 k/N) \approx \sqrt{\ln N}$ to leading order for large $N$ (independently of $k$), 
we see that $\Lambda(\alpha) \to L_N = \sqrt{4D\,\ln (N)/r}$. 
However, the distribution of $M_k$ has exactly the same scaling 
function $f(z) =2\, z\, e^{-z^2}\, \theta(z)$ as 
in (\ref{scaling_kth_max}) except that the scale 
factor $\Lambda(\alpha)$ gets replaced by $L_N$. 
These results are confirmed in our numerical simulations as 
shown in Fig. \newblue{\ref{Fig_combined}a} for different values of $\alpha$. 
Indeed the global maximum $M_1$, in particular, 
typically scales as $L_N\sim \sqrt{\ln N}$ for large $N$. Thus,
even though, on average, the gas is spread over the full real line,
in a typical sample, it is supported over an interval with length
$L_N\sim \sqrt{\ln N}$.
}

\bleu{The behavior of $M_k$ in our correlated gas model is thus very different from the Dyson log-gas or the $1d$ jellium model. In our model, the distributions of the $k$-th maxima, both in and out of the bulk, are described by the same universal scaling function $f(z) = 2 z\, e^{-z^2}\, \theta(z)$. This is in marked contrast to the Dyson log-gas where the distributions of the maxima near the edge are similar to theTracy-Widom distribution while, in the bulk, they are Gaussian \cite{GUS05}. Thus our result for $f(z)$ is a new extreme value \newblue{distribution} that was not encountered before.}

We now turn to the distribution of the spacing (or gap) between two consecutive particles $d_k = M_{k}-M_{k+1}$. 
\bleu{We can exploit again the
renewal structure in Eq. (\ref{Pr}) and write  
\bea \label{pdf_kth_gap}
{\rm Prob.}(d_k = g) = r \int_0^\infty d\tau e^{-r\tau}\, {\rm Prob.}(d_k(\tau) = g) \;,
\eea
where $d_k(\tau) = M_{k}(\tau)-M_{k+1}(\tau)$ is the $k$-th gap of $N$ independent Brownian motions, each of duration $\tau$.  The distribution of the gap $d_k(\tau)$ can be computed in the large $N$ limit, by setting $k = \alpha N$ and using a saddle point method, detailed in~\cite{SM}. We find that $d_k(\tau)$ has a simple exponential distribution 
\bea \label{Pdftau2}
{\rm Prob.}(d_k(\tau) = g)  \approx \frac{b\, N}{\sqrt{\tau}}\,e^{- \frac{b\,N}{\sqrt{\tau}}\,g}  \;,
\eea
where $b = \exp{\left(-[{\rm erfc}^{-1}(2 \alpha)]^2\right)}/\sqrt{4 \pi D}$ is just a constant, independent of $\tau$ and $N$. Inserting this result in Eq. (\ref{pdf_kth_gap}), and performing the change of variable $ u = \sqrt{r\,\tau}$, we obtain
\begin{equation} \label{Pdk_bulk_final}
{\rm Prob.}(d_k = g) \approx \frac{1}{\lambda_N(\alpha)} h \left( \frac{g}{\lambda_N(\alpha)}\right) \,,\, \lambda_N(\alpha) = \frac{1}{b \sqrt{r}N}
\end{equation}
where the normalised scaling function $h(z)$ is given by
\bea\label{h_of_z}
 h(z) = 2 \int_0^\infty du \, e^{-u^2 - \frac{z}{u}} \;.
\eea
The scaling function $h(z) \to \sqrt{\pi}$ as $z \to 0$ and has a 
stretched exponential tail $h(z) \sim e^{-{3}\;(z/2)^{2/3}}$ for large 
$z$ (see \cite{SM}). \newblue{Since $\alpha = k/N$, by choosing $k=1,2,3, \ldots$, one can probe
the first, second, third gap, etc. In this case $\alpha \sim O(1/N)$ is small for large $N$.
We show in \cite{SM} that in this case, $\lambda_N(\alpha) \to \ell_N(k) = 
\sqrt{{D}/{(r\,k^2\,\ln N)}}$.} While the scale factor changes, the 
scaling function $h(z)$ is universal, i.e., independent of $\alpha$. 
This universal result for $h(z)$ is verified in numerical simulations in 
Fig. \newblue{\ref{Fig_combined}b}. From \newblue{Fig. \ref{Fig_combined}b}, it is clear 
that $h(z)$ is a \newblue{monotonically} decreasing function of $z$ with a 
maximum at $z=0$. Thus two consecutive particles are most likely to be 
next to each other (with a zero gap), indicating an effective attraction 
between the particles. This is in stark contrast with the Dyson 
log-gas case where, due to the pairwise repulsion between eigenvalues, 
the spacing distribution vanishes as the gap $g \to 0$: this is the 
celebrated Wigner surmise for the level repulsion in RMT. In addition, 
in the Dyson log-gas as well as in the $1d$ jellium model, the scaling 
functions of the spacing distribution are very different in the bulk and 
at the edges, again in sharp contrast with our result for the 
correlated resetting gas where the gap scaling function $h(z)$ in Eq. 
(\ref{h_of_z}) is universal, i.e., independent of the index $k$ of the 
gap.}

To summarise, we have presented the exact solution of a resetting gas 
with long range correlations in the steady state and computed several 
observables of interest. \bleu{This includes the global average density, 
the distribution of the position of $k$-th rightmost particle and the 
spacing distribution between two consecutive particles. Our technique 
can be easily extended to compute other observables, e.g., the full 
counting statistics, i.e., the distribution of the number of particles 
in a given interval (this is presented in \cite{SM}). Our results
can be generalized to higher dimensions in a straightforward way.} 
Apart from the celebrated log-gas, this is one of the few solvable 
models with strong correlations. In addition, this resetting gas is also 
experimentally realisable. A single diffusing particle with resetting 
has been recently realised in optical trap experiments 
\cite{BBPMC_20,FBPCM_21}, where the particle is allowed to diffuse 
freely for a random time after which a trap is switched on. The particle 
is relaxed to its equilibrium in the \blue{trap} using the "engineering 
swift equilibration" (ESE) technique \cite{ESE_16}. This mimics the 
resetting move of the particle to its equilibrium distribution. The same 
protocol, via ESE, can possibly be implemented to simultaneously 
reset many noninteracting particles in the same optical trap. We thus 
hope that our analytical predictions will stimulate further experimental 
studies of such a resetting gas.

\end{document}


\title{\blue{ Extreme statistics and spacing distribution in a Brownian gas correlated by resetting: Supplemental material }}
\author{Marco Biroli}
\affiliation{LPTMS, CNRS, Univ.  Paris-Sud,  Universit\'e Paris-Saclay,  91405 Orsay,  France}
\author{Hernan Larralde}
\affiliation{Instituto de Ciencias Fisicas, UNAM, Av. Universidad s/n, CP 62210 Cuernavaca Morelos, Mexico}
\author{Satya N. Majumdar}
\affiliation{LPTMS, CNRS, Univ.  Paris-Sud,  Universit\'e Paris-Saclay,  91405 Orsay,  France}
\author{Gr\'egory Schehr}
\affiliation{Sorbonne Universit\'e, Laboratoire de Physique Th\'eorique et Hautes Energies, CNRS UMR 7589, 4 Place Jussieu, 75252 Paris Cedex 05, France}


\begin{abstract}
We give the principal details of the calculations described in the main text of the Letter. 
\end{abstract}

\maketitle

\newpage

\section{Distribution of the $k$-th maximum}

\blue{
As mentioned in the Letter, } 
we consider the resetting gas in the steady state with positions $\{x_1, x_2, \cdots, x_N\}$ ordered as $\{M_1>M_2>\cdots>M_N\}$, where $M_k$ denotes the position of the $k$-th maximum, i.e., the position of the $k$-th particle counted from the right. \blue{As explained in the letter, exploiting the renewal structure of the system we can relate the probability distribution function (PDF) of $M_k$ to the one of the $k$-th maximum of a set of $N$ independent Brownian motions each of duration $\tau$ denoted by $M_k(\tau)$. This is done through equation (7) of the main text which we recall }
\bea \label{pdf_kth_max}
{\rm Prob.}(M_k = w) = r \int_0^\infty d\tau e^{-r\tau}\, {\rm Prob.}(M_k(\tau) = w) \:,
\eea 
\blue{
Hence to study $M_k$ we first need to study $M_k(\tau)$.}

\subsection{\blue{Derivation of extreme value statistics for $M_k(\tau)$}}

\blue{
The maximum of $N$ independent and identically distributed (IID) 
Gaussian variables is well known from the calssical literature of 
extreme value statistics~\cite{Gum_58,david}. we recall this derivation 
here for completeness. As mentioned previously, $M_k(\tau)$ is the 
$k$-th maximum of a set of $N$ independent Brownian motions each of 
duration $\tau$. We recall the position of Brownian motion of duration 
$\tau$ is drawn from the Gaussian distribution
\bea \label{gauss}
p(y, \tau) = \frac{1}{\sqrt{4 \pi D \tau}} e^{-\frac{y^2}{4 D \tau}}.
\eea
}
We first set $k = \alpha N$ and take the large $N$ limit, keeping $\alpha$ fixed. Let us first work out the limiting distribution of the $k$-th maximum $M_k(\tau)$ for fixed $\tau$ and $k = \alpha N$. The PDF of the $k$-th maximum of $N$ IID random variables is given by 
\bea \label{p_k}
{\rm Prob.}(M_k(\tau)=w) = \frac{N!}{(k-1)!(N-k)!} p(w,\tau) \left[ \int_w^\infty p(y,\tau)\, dy\right]^{k-1} \left[\int_{-\infty}^w p(y,\tau)\, dy \right]^{N-k} \;.
\eea
This formula can be understood as follows. Out of $N$ IID variables, we fix the value of the $k$-the maximum to be $w$, then there are $(k-1)$ variables above $w$ and $(N-k)$ variables below $w$. Using the independence of the variables and taking into account the number of ways of arranging this ordering (this is encoded in the combinatorial factor in Eq.~(\ref{p_k})), one arrives at Eq. (\ref{p_k}).
We now set $k = \alpha N$ and rewrite this as
\bea \label{p_k_2}
{\rm Prob.}(M_k(\tau)=w) = \frac{N!}{\Gamma(\alpha N)\Gamma[(1-\alpha)N+1]}  \frac{p(w,\tau)}{\int_w^\infty p(y,\tau)dy} e^{- N \Phi_\alpha(w)}
\eea
where 
\bea \label{Phi_a}
\Phi_\alpha(w)  = - \alpha \ln \left( \int_w^\infty p(y,\tau) dy\right) - (1-\alpha) \ln \left( \int_{-\infty}^w p(y,\tau)dy \right) \;.
\eea
Note that, for convenience, we have expressed the factorials in terms of Gamma functions in Eq. (\ref{p_k_2}). This formula~(\ref{p_k_2}) is exact for all $N$. As $N \to \infty$, this PDF gets sharply peaked around the minimum of $\Phi_\alpha(w)$, say at $w^*$. The location of this minimum $w^*$ can be easily computed by minimising $\Phi_\alpha(w)$. Setting $\Phi_\alpha'(w=w^*) = 0$ one immediately gets
\bea \label{wstar}
\int_{w^*}^\infty p(y,\tau) dy = \alpha \;.
\eea
Note that $w^*$ is called the $\alpha$-quantile as it denotes the location of $y$ above which the average fraction of particles is $\alpha$.
In our case $p(y,\tau)$ is a Gaussian distribution given in (\ref{gauss}) and this relation (\ref{wstar}) reads explicitly
\bea \label{w*_gauss}
w^* = \sqrt{4 D \,\tau}\, {\rm erfc}^{-1}(2 \alpha) \;,
\eea
where ${\rm erfc}(z) = \frac{2}{\sqrt{\pi}}\int_z^\infty e^{-u^2}\,du$ is the complementary error function and ${\rm erfc}^{-1}(z)$ is its inverse. Expanding $\Phi_\alpha(w)$ around $w = w^*$ up to quadratic order, one finds after straightforward algebra that around $w = w^*$, and for large $N$, the PDF of $M_k(\tau)$ takes a Gaussian form
\bea \label{M_k_gauss}
{\rm Prob.}(M_k(\tau) = w) \approx \sqrt{\frac{N}{2 \pi \alpha(1-\alpha)}} \, p(w^*,\tau) \exp\left(- \frac{N\,p^2(w^*,\tau)}{2\alpha(1-\alpha)}(w-w^*)^2 \right) \;.
\eea
This is a normalised Gaussian distribution centered around $w^*$ and with a width that decays as $1/\sqrt{N}$ for large $N$. Indeed, for large $N$, it essentially approaches a delta-function, centered at $w = w^*$ given in Eq. (\ref{w*_gauss}). 

\blue{Now, substituting this limiting Gaussian distribution 
in Eq. (\ref{pdf_kth_max}) it is easy to see that to leading order for 
large $N$, one can ignore the fluctuations of $M_k(\tau)$ around its mean 
$w^*$ (since its
variance decays as $1/N$) and just replace the Gaussian by a delta function
centered at $w^*$}, 
\bea \label{delta_w*}
{\rm Prob.}(M_k(\tau) = w)  \approx \delta(w - w^*) = \delta\left(w- \sqrt{4 D\,\tau}\, {\rm erfc}^{-1}(2 \alpha)\right)  \;. 
\eea
Performing the resulting integral over $\tau$ trivially, we get for the PDF of $M_k$
\bea \label{P_Mk_alpha}
P(M_k) \approx \frac{1}{\Lambda(\alpha)} f\left( \frac{M_k}{\Lambda(\alpha)}\right) \quad \quad {\rm with} \quad \quad \Lambda(\alpha) =  \sqrt{\frac{4D}{r}} {\rm erfc}^{-1}(2 \alpha)\;,
\eea
where $f(z)  = 2\,z\,e^{-z^2}$ \blue{for $z \geq 0$} is the \blue{normalized} scaling function \blue{given in the main text in Eq. (8).}

\subsection{\blue{Limiting ($\alpha \to 0$) behavior.}}

\blue{
From Eq. (\ref{P_Mk_alpha}) we see that the scaling function 
$f(z) = 2\, z\, e^{-z^2}\, \theta(z)$ is completely independent 
of $\alpha$, only the scale factor $\Lambda(\alpha)$ depends on $\alpha$. 
In order to probe the behavior of the gas close to the first maxima, 
i.e., $k = \mathcal{O}(1)$ we then have to take the $\alpha = k/N \ll 1$ 
limit. To do so, note that ${\rm erfc}^{-1}(2\alpha)$ becomes very large 
as $\alpha \to 0$, hence we can use the well known large-$z$ asymptotic 
of ${\rm erfc(z)} \sim e^{-z^2}/(z\sqrt{\pi})$ to write
\bea \label{eq:erfc_asymp}
2 \alpha = {\rm erfc}\left( {\rm erfc}^{-1}\left[2 \alpha\right] \right) = \frac{e^{-[{\rm erfc}^{-1}(2\alpha)]^2}}{{\rm erfc}^{-1}(2\alpha) \sqrt{\pi}} .
\eea
Hence to leading order, when $\alpha = k/N$ with $k \sim \mathcal{O}(1)$ and $N \gg 1$, we have that 
\bea \label{eq:a2N}
{\rm erfc}^{-1}(2\alpha) = \sqrt{-\ln(2\alpha)} = \sqrt{\ln (N/(2k))} \sim \sqrt{\ln N},
\eea
and plugging this back in Eq. (\ref{P_Mk_alpha}) we get
\bea \label{eq:Ln}
\Lambda(\alpha) \stackrel{\alpha = k/N}{\longrightarrow} L_N = \sqrt{\frac{4 D \ln N}{r}}.
\eea
Note that close to the global maximum the scale factor, 
which is now given by $L_N$, becomes completely independent of $k$ to
leading order for large $N$. Hence the whole distribution 
$P(M_k)$ is identical for particles close to the global maximum. 
Strikingly, as pointed out in the main text, we see that the behavior of our gas is universal in and out of the bulk. The scaling function which determines the successive positions of the maxima is everywhere the same and only the scale factor changes, smoothly crossing over from $\Lambda(\alpha)$ in the bulk to $L_N$ close to the global maximum.
}

\subsection{\blue{Check for $k = \mathcal{O}(1)$}}

\blue{In proving the universality of $f(z)$ above, we have
extrapolated our bulk calculation, performed above with $k=\alpha\, N$
with $\alpha$ fixed and $N$ large, to the case
when $\alpha\sim O(1/N)$. In this subsection, we 
present an alternative derivation specific to the case $k = 
\mathcal{O}(1)$ and show that indeed we reach the same result, i.e.,
his extrapolation is fully justified.}

It is well known from the EVS of IID random variables that in the large $N$ limit, the PDF of $M_k(\tau)$, for fixed $k = O(1)$, converges to the following distribution \cite{david,SM_14,MPS_20}
\bea \label{gen_gumbel}
{\rm Prob.}(M_k(\tau) = w) \to \frac{1}{b_N(\tau)}\, G_k \left( \frac{w-a_N(\tau)}{b_N(\tau)}\right) \;,
\eea
where 
\bea 
a_N(\tau) \approx \sqrt{4 D \,\tau \ln  {N} } \quad\quad, \quad \quad b_N(\tau)  \approx \sqrt{\frac{D\,\tau}{\ln N}} \label{aNbN} 
\eea
and $G_k(z)$ is the generalized Gumbel PDF
\bea \label{gen_gumbel2}
G_k(z) = \frac{1}{(k-1)!} e^{-kz - e^{-z}} \;.
\eea
We now insert this scaling form (\ref{gen_gumbel}) in the integral in Eq. (\ref{pdf_kth_max}) and perform the change of variable $\tau \to z$ with 
\bea \label{change}
\frac{w-a_N(\tau)}{b_N(\tau)} = z \quad \Longrightarrow \quad \tau = \frac{\ln N}{D} \frac{w^2}{\left(z + 2 \ln N\right)^2} \;.
\eea
Taking the scaling limit $N \to \infty$, $w \to \infty$ but keeping $w/\sqrt{\ln N}$ fixed and using the normalization $\int_{-\infty}^\infty G_k(z)dz=1$, one gets from Eq. (\ref{pdf_kth_max}) 
\bea \label{largeN_kth_max}
{\rm Prob.}(M_k=w) \approx \frac{r\,w}{2D\,\ln N} \exp\left(- \frac{r\,w^2}{4 D \,\ln N } \right) \;.
\eea
Then the PDF $P(M_k)$ of the $k$-th maximum can be written in the scaling form 
\bea \label{scaling_kth_max}
P(M_k) \approx \frac{1}{L_N} \, f\left( \frac{M_k}{L_N}\right) \quad\quad {\rm with} \quad \quad L_N = \sqrt{\frac{4D\, \ln N}{r}}
\eea 
and the scaling function $f(z)$ is given by
\bea \label{scaling_f}
f(z) = 2 z \, e^{-z^2} \quad, \quad z \geq 0 \;.
\eea
\blue{
Thus we recover the exact same result as the one derived previously 
in Eq. (\ref{eq:Ln}), justifying the extrapolation
of the bul result with $\alpha\sim O(1)$ to the
case when $\alpha\sim O(1/N)$.
}

\section{Distribution of the $k$-th gap}

\blue{
Here we consider the behavior of the $k$-th gap $d_k = M_k - M_{k+1}$. 
As explained in the Letter, by exploiting the renewal structure of the 
system we can relate the PDF of $d_k$ to the PDF of the 
$k$-th gap, $d_k(\tau) = M_k(\tau) - M_{k+1}(\tau)$,
of a set of $N$ independent Brownian motions each of 
duration $\tau$. This is demonstrated in Eq. (9) of the 
main text which we recall}
\bea \label{kth-gap}
{\rm Prob.}(d_k = g) = r \int_0^\infty d\tau \,e^{-r\tau}\, {\rm Prob.}\left[M_k(\tau) - M_{k+1}(\tau) = g \right]  =  r \int_0^\infty d\tau \,e^{-r\tau}\, {\rm Prob.}\left[d_k(\tau) = g \right] \;,
\eea 
where $M_{k}(\tau)$, as before, is the $k$-th maximum of $N$ IID random variables, each distributed via the Gaussian distribution $p(y,\tau)$, parametrised by $\tau$, in Eq. (\ref{gauss}). Thus we need to first find the distribution of the gap $d_k(\tau) = M_k(\tau) - M_{k+1}(\tau)$ for fixed $\tau$. 

\blue{
\subsection{Derivation of the PDF of $d_k(\tau)$}
}
To study the behavior of $d_k(\tau)$ we need to start with the joint distribution of $M_k(\tau)$ and $M_{k+1}(\tau)$ for $N$ IID \blue{Gaussian} random variables drawn from $p(y, \tau)$, parametrised by $\tau$ and defined in Eq. (\ref{gauss}). Then
\bea \label{joint_Mk}
{\rm Prob.}\left[ M_k(\tau) = x, M_{k+1}(\tau)=y\right] = \frac{N!}{(k-1)! (N-k-1)!} p(x,\tau)p(y,\tau)\left[\int_x^\infty p(x',\tau) dx'\right]^{k-1}\left[ \int_{-\infty}^y p(x',\tau) dx'\right]^{N-k-1} 
\eea
for $x>y$. The interpretation is again simple, as in Eq. (\ref{p_k}). We choose two out of $N$ variables and fix their positions at $x$ and $y<x$. There are $(k-1)$ variables above $x$ and $N-k-1$ variables below $y$. The combinatorial factor just counts the number of ways of ordering. \\\\
We now set $k = \alpha N$ and rewrite Eq. (\ref{joint_Mk}) as 
\bea \label{gap_bulk1}
{\rm Prob.}\left[ M_k(\tau) = x, M_{k+1}(\tau)=y\right]  =\frac{\Gamma(N+1)}{\Gamma(\alpha N) \Gamma[(1-\alpha)N]}\; U(x,\tau)\, V(y,\tau)\, e^{N S_\alpha(x,y)} \;,
\eea
where
\bea \label{UV}
U(x,\tau) =  \frac{p(x,\tau)}{\int_x^\infty p(x',\tau) dx'} \quad, \quad \quad V(y,\tau) = \frac{p(y,\tau)}{\int_{-\infty}^y p(y',\tau) dy'} \;,
\eea
and $S_\alpha(x,y)$ reads
\bea \label{Salpha}
S_\alpha(x,y) = \alpha \ln \left[ \int_x^\infty p(x',\tau)dx'\right] + (1-\alpha) \ln\left[ \int_{-\infty}^y p(x',\tau)\,dx'\right] \;.
\eea
From this joint distribution (\ref{gap_bulk1}) one can compute the gap distribution by setting $x  = y + g$ and integrating over $y$ with $g \geq 0$ fixed. This gives
\bea \label{Pdktau}
{\rm Prob.}(d_k(\tau) = g) = \frac{\Gamma(N+1)}{\Gamma(\alpha N) \Gamma[(1-\alpha)N]}\; \int_{-\infty}^\infty dy \, U(y+g,\tau)\, V(y,\tau)\, e^{N S_\alpha(y+g,y)} \;.
\eea
In the large $N$ limit, this form suggests to evaluate the integral over $y$ using a saddle point method. The saddle point is attained at $y=y^*$ where $\partial S/\partial y |_{y=y^*} = 0$. We expect that, in the bulk, the typical gap scales as $O(1/N)$ and hence is small. Therefore one can find the solution of the saddle point equation $y^*$ in powers of $g$ and one finds that 
\bea \label{y*}
y^* = w^*  + A\,g + O(g^2) \quad \quad {\rm with} \quad \quad \int_{w^*}^\infty p(y,\tau) dy  =\alpha \;.
\eea
where $A$ is a computable constant, whose actual value turns out to be irrelevant to leading order in the large $N$ limit. In our case $w^*$ is given exactly in Eq. (\ref{w*_gauss}). Evaluating the saddle-point action at $y=y^*$ one gets, 
\bea \label{S_sp}
S(y^*+ g,y^*) = \alpha \ln\alpha + (1-\alpha)\ln(1 - \alpha) - p(w^*,\tau)\,g + O(g^2) \;.
\eea
The first two terms in Eq. (\ref{S_sp}) cancel exactly the combinatorial factor in (\ref{Pdktau}), expanded using Stirling's formula for large $N$. 
Evaluating this integral over $y$ by the saddle point method and carefully collecting all the factors, we find, after a bit of algebra, a rather simple
expression, namely 
\bea \label{Pdftau2}
{\rm Prob.}(d_k(\tau) = g)  \approx N p(w^*,\tau)\, e^{- N\,p(w^*,\tau)\,g} \quad \quad {\rm where} \quad \quad w^* = \sqrt{4 D \,\tau}\, {\rm erfc}^{-1}(2 \alpha) 
\eea
and $p(w^*,\tau)$ reads explicitly
\blue{
\bea\label{pystar}
p(w^*,\tau) = \frac{1}{\sqrt{4 \pi D \,\tau}}\, 
\exp\left(-[{\rm erfc}^{-1}(2\alpha)]^2\right) \; .
\eea
For simplicity, let us introduce
\bea \label{eq:b}
p(w^\star, \tau) = \frac{b}{\sqrt{\tau}} \quad \quad {\rm where} \quad \quad b = \frac{ \exp(-[{\rm erfc}^{-1}(2\alpha)]^2)}{\sqrt{4 \pi D}} \;.
\eea
}
Finally, substituting this scaling form (\ref{Pdftau2}) in Eq. (\ref{kth-gap}), one finds after a simple change of variable ($u = \sqrt{r \tau}$) that the PDF of the $k$-th gap with $k = \alpha N$ can be expressed in the scaling form [see Eq. (11) in the main text]
\blue{
\bea \label{p_gap_bulk_final}
{\rm Prob.}(d_k = g) \approx \frac{1}{\lambda_N(\alpha)}\,
h\left( \frac{g}{\lambda_N(\alpha)}\right) \quad \quad {\rm with} \quad \quad \lambda_N(\alpha) = \frac{1}{b \sqrt{r} N} \;,
\eea
where the normalized scaling function $h(z)$ defined for $z \geq 0$ is given by
\bea \label{eq:hz1}
h(z) = 2 \int_0^{+\infty} du \; e^{-u^2 - \frac{z}{u}}.
\eea
Hence recovering the result given in Eq. (12) of the main text.
}

\blue{
\subsection{Limiting $(\alpha \to 0)$ behavior}
From Eq. (\ref{p_gap_bulk_final}) we see that the scaling function 
$h(z)$ is completely universal, i.e., independent of $\alpha$. 
Only the scale factor $\lambda_N(\alpha)$ depends on $\alpha$.
This is similar to what happened for the $k$-maximum $M_k$ before. 
Now to probe the behavior close to the global maximum, i.e., 
when $k = \mathcal{O}(1)$ and $\alpha = k/N \ll 1$ we use the asymptotic we previously derived for $M_k$ in Eq. (\ref{eq:erfc_asymp}). Replacing the result in the expression for $b$ given in Eq. (\ref{eq:b}) we obtain
\bea
b = \frac{ \exp(-[{\rm erfc}^{-1}(2\alpha)]^2)}{\sqrt{4 \pi D}} \stackrel{\alpha \ll 1}{\longrightarrow} \frac{2 \alpha \, {\rm erfc}^{-1}(2\alpha)}{\sqrt{4 D}} \;.
\eea
Now replacing with $\alpha = k/N$ and using Eq. (\ref{eq:a2N}) we get
\bea
b \stackrel{\alpha = k / N}{\longrightarrow} \frac{2 k \sqrt{\ln N}}{N \sqrt{4 D}} \;,
\eea
and placing this result back in Eq. (\ref{p_gap_bulk_final}) we get
\bea \label{eq:ell}
\lambda_N(\alpha) \stackrel{\alpha = k/N}{\longrightarrow} \ell_N(k) = \sqrt{\frac{D}{r k^2 \ln N}} \;.
\eea
Notice that, unlike $M_k$, the scale factor is still dependent on $k$ close to the global maximum. However, as for $M_k$, the behavior of the gap is universal in and out of the bulk, i.e. the scaling function never changes. Only the scale factor changes smoothly crossing over from $\lambda_N(\alpha)$ in the bulk to $\ell_N(k)$ close to the global maximum.
}

\blue{\subsection{Check for $k = O(1)$}
As in the case of $M_k$, we can also derive the result for $d_k$
when $k\sim O(1)$ independently.}
The case $k=O(1)$ has been studied extensively in the literature for IID random variables \cite{SM_14,MPS_20} and one can show, starting from Eq. (\ref{joint_Mk}) that, in the limit $N \to \infty$ with $k$ fixed, 
\bea \label{rel_dk_sk}
d_k(\tau) \to b_N(\tau) \,s_k \;,
\eea
where $b_N(\tau) = \sqrt{D\,\tau/\ln N}$ and $s_k$ is a random variable (independent of $N$) distributed via the exponential law
\bea \label{exp_sk}
{\rm Prob.}(s_k=s) = k \, e^{-k\,s} \;, \; s \geq 0 \;.
\eea 
By substituting the result (\ref{rel_dk_sk}) in (\ref{kth-gap}) and performing a simple change of variable, one arrives at the PDF of the $k$-th gap when $k=O(1)$ 
\begin{equation}\label{gap_sm}
P(d_k) \approx \frac{1}{\ell_N(k)} h\left( \frac{d_k}{\ell_N(k)}\right) \quad \quad {\rm with} \quad \quad \ell_N(k) = \sqrt{\frac{D}{r\,k^2\,\ln N}} \;.
\end{equation}
\blue{Hence we recover the same scale factor as derived in Eq. (\ref{eq:ell}) and the same scaling function}
\bea \label{scaling_h}
h(z) = 2 \int_0^\infty du \, e^{-u^2 - \frac{z}{u}}  \;,
\eea 
\blue{as derived previously in Eq. (\ref{eq:hz1})}.

\subsection{Asymptotic behavior of the scaling function $h(z)$}
We derive the asymptotic behaviors of the scaling function $h(z)$ given in Eq. (\ref{eq:hz1}), namely
\bea \label{hz2}
h(z) = 2 \int_0^\infty du \, e^{-u^2 - \frac{z}{u}}  \;.
\eea
For $z \to 0$, one trivially has $h(0) = \sqrt{\pi}$. However, one sees that the function is not analytic near $z =0$ since a naive Taylor expansion of the integrand in powers of $z$ yields diverging integrals. One can actually split the integral into two intervals $[0,z]$ and $[z, +\infty)$. The contribution from the first interval is linear in $z$ for small $z$. The leading singular correction comes from the second interval where we can expand $e^{-z/u}$ in powers of $z$. The first term gives $\sqrt{\pi}$ as $z \to 0$, the second term behaves as $-2 z \int_z^{\infty} e^{-u^2}/u$, which to leading order for small $z$ behaves as $2 z \ln z$. Hence, for small $z$ we get
\bea \label{h_small}
h(z) = \sqrt{\pi}  + 2 z \, \ln z + O(z) \;. 
\eea
The large $z$ behavior can be obtained easily by a standard saddle point method (we do not provide details here). In summary, the asymptotic behaviors of $h(z)$ are given by
\bea \label{asympt_h}
h(z) \approx
\begin{cases}
&\sqrt{\pi} + 2 z \ln z \quad, \quad \hspace*{1.5cm} z \to 0 \\
& \\
&2\sqrt{\frac{\pi}{3}} \exp{\left(-3 \left( \frac{z}{2}\right)^{2/3} \right)} \quad, \quad z \to \infty \;.
\end{cases}
\eea 
Thus the universal scaling function $h(z)$ has rather nontrivial asymptotic behaviors. Its derivatives diverges logarithmically at $z = 0$ and it has
a stretched exponential tail for large $z$, with a stretching exponent $2/3$. A plot of this function is shown in Fig. 2b in the main text. \\\\

\section{Numerical simulations}

We briefly outline here the method of numerical simulations used in the main text. We consider $N$ Brownian particles on a line, each with the 
same diffusion constant $D$. They all start at the origin at $t=0$.  Let $x_i(t)$ denote the position of the $i$-th particle at time $t$. These positions evolve by the following stochastic rule. In a small time $\Delta t$
\bea \label{discrete1}
x_i(t+\Delta t) = 
\begin{cases}
&0 \quad \hspace*{3cm}{\rm with \; prob.} \; r \Delta t \;, \\
&\\
& x_i(t) + \sqrt{2 D \,\Delta t}\, \eta_i(t) \quad {\rm with \; prob.} \; 1 - r \Delta t
\end{cases}
\eea
where $r$ is the resetting rate and $\eta_i(t)$ are IID Gaussian random variables with zero mean and unit variance. Note that this equation holds for all $i$, in particular the first line in Eq. (\ref{discrete1}) shows that when a resetting event happens, the particles are all simultaneously reset to the origin. This gives us the trajectories of the gas of $N$ particles, i.e., the vector $\{x_1(t), x_2(t), \cdots, x_N(t) \}$ at all time $t$. In the long time limit, the distribution of $\{x_1(t), x_2(t), \cdots, x_N(t) \}$ approaches a non-equilibrium stationary state $P_{\rm stat}[\{ x_i\}]$ as given in Eq. (3) of the main text. Numerically, we keep track of this trajectory vector $\{x_1(t), x_2(t), \cdots, x_N(t) \}$ and measure different observables from it in the stationary state. The results presented in the main text in Fig. 2 are then obtained by averaging over $10^5$ samples. We used the parameter values $D=0.5$ and $r=1$. 

\blue{
\section{Full counting statistics}
So far we have presented exact results in the corelated resetting gas
in its stationary state for three basic observables
namely the global average density, the $k$-th maximum $M_k$ and
the $k$-th gap $d_k$. In fact, our
method can be easily generalized to compute other observables, 
such as the full counting statistics (FCS), i.e., the distribution 
$P(N_L,N)$ of the number of particles $N_L$ contained in a 
symmetric interval $[-L,+L]$ around the resetting position $x=0$. Once more, exploiting the renewal structure of the system, as done in the main text, we can write it as
\begin{equation} \label{FCS}
P(N_L,N) = r \int_0^\infty \hspace*{-0.2cm}d\tau\, e^{-r \tau} \, {N \choose N_L} \, \left[q(\tau)\right]^{N_L} \left[1 - q(\tau)\right]^{N-N_L} \hspace*{-0.3cm} .
\end{equation}
Here $q(\tau) = \int_{-L}^L p(y,\tau)\, dy = {\rm erf}(L/\sqrt{4 D \tau})$ with ${\rm erf}(z) = 1 - {\rm erfc}(z)$ which denotes the probability that a single Brownian particle, at time $\tau$, is inside the interval $[-L,+L]$. The binomial distribution inside the integrand just denotes the probability that $N_L$, out of $N$ independent particles, are in the interval $[-L,+L]$ at time $\tau$. Setting $N_L = \kappa N$, with $0<\kappa<1$ fixed, the binomial distribution converges to a Gaussian distribution with mean $N\,q(\tau)$ and variance $N q(\tau)(1-q(\tau))$. As in the case of the maximum $M_k$, the fluctuations of this Gaussian variable do not contribute to the integral in the large $N$ limit and one can replace the Gaussian by a delta-function $\delta(N_L - N\,q(\tau))$ leading to
\bea \label{FCS2}
P(N_L,N) \approx \frac{r}{N} \int_0^\infty d\tau\, e^{-r \tau}  \delta(\kappa - q(\tau)) \;.
\eea
Using the explicit form of $q(\tau)$, the integral over $\tau$ can now be performed by a change of variable and $P(N_L,N)$ takes the scaling form
\bea \label{FCS_scaling}
P(N_L,N) \approx \frac{1}{N} \, H\left(\frac{N_L}{N} \right) \;,
\eea
where the scaling function $H(\kappa)$ (with $0\le \kappa\le 1$) is given by 
\bea \label{H}
H(\kappa) = \gamma \sqrt{\pi} 
\left[u(\kappa)\right]^{-3}\, 
\exp\left[-\frac{\gamma}{u(\kappa)^2} + [u(\kappa)]^2\right]\, .
\eea
Here $\gamma = rL^2/(4D)$ and $u(\kappa) = {\rm erf}^{-1}(\kappa)$. The PDF $H(\kappa)$ of $0\leq\kappa\leq 1$ is normalised to unity $\int_0^1 H(\kappa)\, d\kappa = 1$ and has an unusual non-trivial shape [see Fig. \ref{Fig_fcs}]. As $\kappa \to 0$, the function $H(\kappa) \approx\frac{8\,\gamma}{\pi \kappa^3} \exp\left(-\frac{4 \gamma}{\pi \kappa^2} \right)$ vanishes very fast, while it diverges (though still integrable) as 
$H(\kappa) \approx \frac{\gamma}{(1-\kappa) |\ln (1-\kappa)|^2}$
as $\kappa \to 1$.
Numerical simulations are in very good agreement with our analytical prediction in Eq. (\ref{H}). Thus the scaling form of the FCS in Eqs. (\ref{FCS_scaling})-(\ref{H}) is fundamentally different from the log-gas case. Here the mean and standard deviation of $N_L$ both scale as $N$, while in the log-gas the mean scales as $N$ and the Gaussian fluctuations around the mean have 
standard deviation $\sim \sqrt{\ln N}$~\cite{FS_95,CL_95,MNSV_09,MMSV_14,SLMS_21}. 
\begin{figure}
\centering
\includegraphics[width = 0.5 \linewidth]{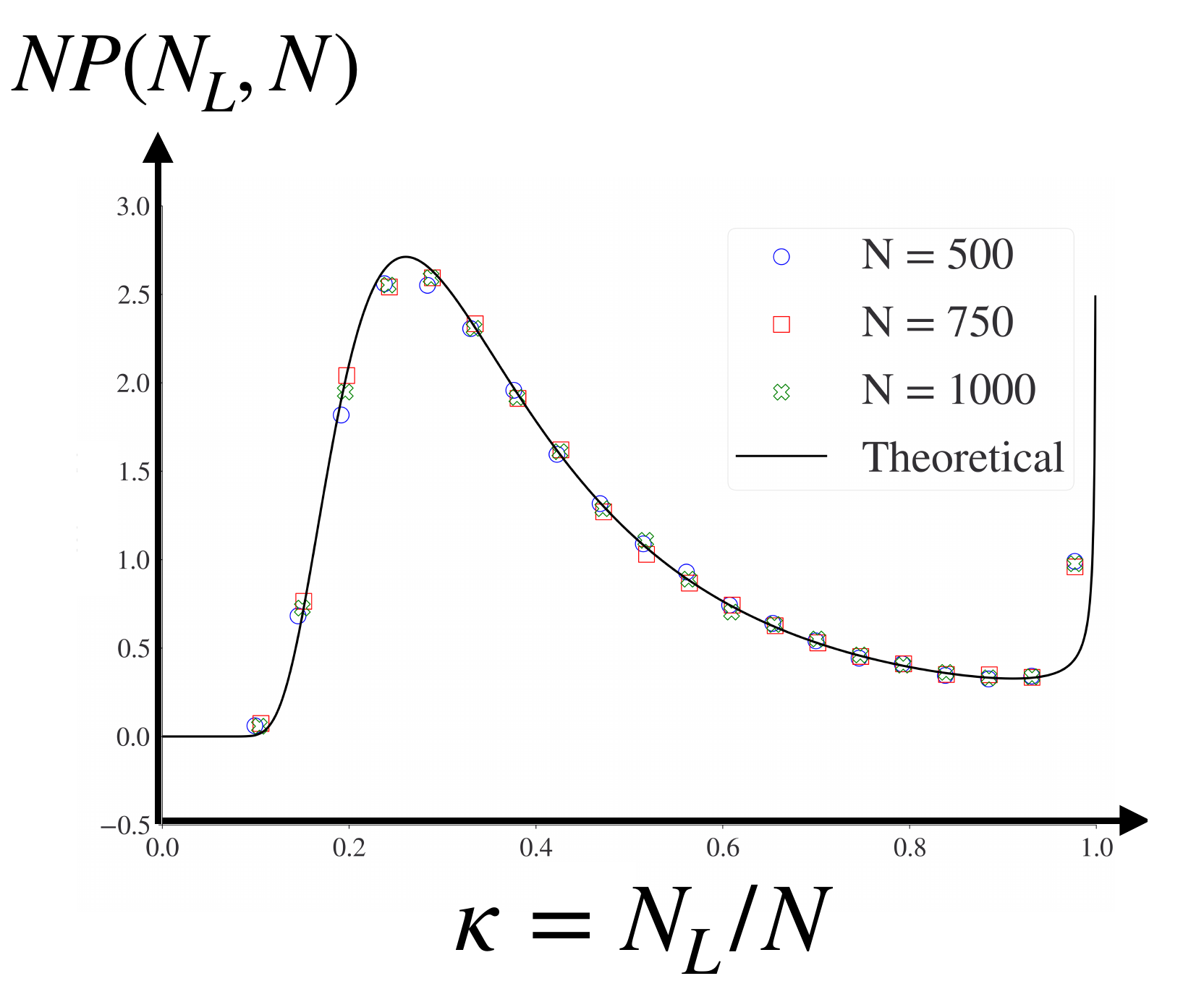}
\caption{\label{Fig_fcs} Numerical results for FCS in $[-L, +L]$ (with $L = 0.4$) compared with the analytical predictions in Eqs. (\ref{FCS_scaling}) and (\ref{H}). We used the parameter values $D = 0.5$ and $r = 1$.}
\end{figure}
}